\documentclass[aps,prl,twocolumn,groupedaddress]{revtex4-1}
\usepackage{graphicx,epsfig,dcolumn,afterpage}
\usepackage{colortbl}
\usepackage[latin1]{inputenc} 
\usepackage{amsmath}

\usepackage{color}
\usepackage{colortbl}

\newcommand{\TV}[1]{}  

\begin{document}

\title{\emph{Ab Initio} Simulation of Electrical Currents Induced by Ultrafast Laser Excitation of Dielectric Materials}

\author{Georg Wachter$^1$}
\email{georg.wachter@tuwien.ac.at}
\author{Christoph Lemell$^1$}
\author{Joachim Burgd\"orfer$^1$}
\affiliation{$^1$Institute for Theoretical Physics, Vienna University of Technology, 1040 Vienna, Austria, EU}

\author{Shunsuke A. Sato$^2$}
\author{Xiao-Min Tong$^{2,3}$}
\author{Kazuhiro Yabana$^{2,3}$}
\affiliation{$^2$Graduate School of Pure and Applied Sciences, University of Tsukuba, Tsukuba 305-8571, Japan}
\affiliation{$^3$Center for Computational Sciences, University of Tsukuba, Tsukuba 305-8577, Japan}

\date{\today}

\begin{abstract}
We theoretically investigate the generation of ultrafast currents in insulators induced by strong few-cycle laser pulses. \emph{Ab initio} simulations based on time-dependent density functional theory give insight into the atomic-scale properties of the induced current signifying a femtosecond-scale insulator-metal transition. We observe the transition from nonlinear polarization currents during the laser pulse at low intensities to tunnelinglike excitation into the conduction band at higher laser intensities. At high intensities, the current persists after the conclusion of the laser pulse considered to be the precursor of the dielectric breakdown on the femtosecond scale. We show that the transferred charge sensitively depends on the orientation of the polarization axis relative to the crystal axis suggesting that the induced charge separation reflects the anisotropic electronic structure. We find good agreement with very recent experimental data on the intensity and carrier-envelope phase dependence [A.~Schiffrin et al., Nature (London) 493, 70 (2013)]. 
\end{abstract}


\maketitle

The availability of femtosecond laser sources providing wave-form controlled high-intensity pulses has opened up novel opportunities to explore the ultrafast and nonlinear response of matter. While experiments with rare-gas targets have provided considerable insight into the real-time motion of electrons within atoms and the nonlinear optical response in terms of high-harmonic generation \cite{Krausz2009Attosecond}, the exploration of laser-induced subfemtosecond processes in the realm of solid-state and surface physics is only at the very beginning \cite{Cavalieri2007Attosecond,Gertsvolf2008Orientationdependent,*Gertsvolf2010Demonstration,Ghimire2011Observation}. Delicate light-field control of electron currents emitted from surfaces, nanostructures, and nanoparticles has been demonstrated \cite{Lemell2003Electron,Dombi2004Direct,Kruger2011Attosecond,Wachter2012Electron,Zherebtsov2011Controlled}, while experiments with bulk matter have succeeded to monitor the electronic dynamics indirectly through optical signals \cite{Mitrofanov2011Optical}.

Very recently, Schiffrin \emph{et al.} \cite{Schiffrin2013Opticalfieldinduced} have shown that strong few-cycle laser pulses induce currents and charge separation in large band-gap dielectrics. In contrast to the electrical current and subsequent electron-avalanche breakdown induced by static fields or picosecond laser pulses \cite{Sparks1981Theory}, the charge transfer observed in Ref.~\cite{Schiffrin2013Opticalfieldinduced} results from optical-field-induced transient and reversible currents below the destruction threshold. These results suggest that the intense laser field strongly distorts the electronic band structure thereby converting an insulator transiently into a metal on the (sub) femtosecond scale. This picture is supported by first modeling efforts based on independent-particle models \cite{Schiffrin2013Opticalfieldinduced,Foldi2013Effect,Korbman2013Quantum,Hawkins2013Role} including macroscopic screening effects \cite{Kruchinin2013Theory,Apalkov2012Theory,Schiffrin2013Opticalfieldinduced}. Since previous approaches are based on mostly one-dimensional phenomenological models, the interrelation between the ultrafast dynamics and the microscopic lattice and electronic structure of the dielectric remains to be understood. Open questions include the following: Does the current correspond to a nonlinear Maxwellian polarization current leading to a finite polarization after the end of the driving laser pulse? Is the current due to electrons nonadiabatically transferred into the conduction band where transport might continue after the conclusion of the driving laser pulse? Does the relative orientation of the laser polarization and crystallographic axes influence the ultrafast response? 


In this Letter, we investigate the origin of optical field-induced currents in bulk insulators on the atomic scale. We present the first fully three-dimensional \emph{ab initio} simulations based on time-dependent density functional theory (TDDFT). Our simulations give access to the dynamics of microscopic quantities including the spatially and time-resolved electron density and electrical current density within the unit cell of the material. The simulations thereby provide unprecedented insight into the spatiotemporal structure of the charge dynamics on the atomic length and time scale. 

We employ a real-space, real-time formulation of TDDFT \cite{Yabana1996Timedependent,Bertsch2000Realspace,Onida2002Electronic,Marques2004Timedependent,Otobe2009Firstprinciples} to simulate the electronic dynamics induced by strong few-cycle laser pulses in $\alpha$-SiO$_2$ ($\alpha$-quartz). 
Details of the simulation for $\alpha$-SiO$_2$ have been reported in Ref.~\cite{Otobe2009Firstprinciples}. Briefly, we solve the time-dependent Kohn-Sham equations (atomic units used unless stated otherwise)
\begin{equation}
  i \partial_t \psi_i(\mathbf{r},t) = H(\mathbf{r},t)\psi_i(\mathbf{r},t) \quad , 
\label{eq:tdks}
\end{equation}
where the index $i$ runs over the occupied Kohn-Sham orbitals $\psi_i$ with the Hamiltonian
\begin{equation} \label{eq:hamiltonian}
H(\mathbf{r},t) = 
\frac{1}{2} \left( -i\mathbf{\nabla} + \mathbf{A}(t) \right)^2 
+ \hat{V}_\mathrm{ion} 
+ \int d\mathbf{r'} \frac{ n(\mathbf{r'},t) }{ | \mathbf{r}-\mathbf{r'} | } + \hat{V}_\mathrm{XC}(\mathbf{r},t)
\end{equation}
describing the system under the influence of a homogenous time-dependent electric field $\mathbf F(t)$ of amplitude $\mathbf F_0$ with a vector potential $\mathbf A(t) = - \int_{-\infty}^t \mathbf F(t') dt' $ in the velocity gauge and in the transverse geometry \cite{Yabana2012Timedependent}. While in the so-called longitudinal geometry \cite{Yabana2012Timedependent}, charging up of dielectric interfaces at finite distances is included, the transverse geometry allows us to treat the bulk polarization response of the infinitely extended system along the polarization direction. The latter appears to be a reasonable approximation mimicking the presence of metallic contacts not explicitly treated in the simulation. The periodic lattice potential $\hat{V}_\mathrm{ion}$ is given by norm-conserving pseudopotentials of the Troullier-Martins form \cite{Troullier1991Efficient} representing the ionic cores (O(1s$^2$) and Si(1s$^2$2s$^2$2p$^6$)). The valence electron density is given as $n(\mathbf{r},t) = \sum_i |\psi_i(\mathbf{r},t)|^2$. 
For the exchange and correlation (XC) potential $\hat{V}_\mathrm{XC}$, we employ the adiabatic Tran-Blaha modified Becke-Johnson meta-GGA functional \cite{Tran2009accurate,*Koller2011Merits,*Koller2012improving,Nazarov2011Optics}. It accurately reproduces the band gap in ground-state calculations yielding $\sim 9$ eV for SiO$_2$ and yields good agreement with the experimental dielectric function over the entire spectral range of interest including at optical frequencies \cite{Philipp1966Optical}.
Since the coupling to the time-dependent external field enters in terms of the vector potential, the Hamiltonian (Eq. \ref{eq:hamiltonian}) can be alternatively viewed as the starting point of time-dependent \emph{current} density functional theory \cite{Vignale1996Currentdependent,Vignale1997Timedependent} in the adiabatic approximation, in which the nonadiabatic exchange-correlation contribution to the vector potential $\mathbf{A}_{\mathrm{XC}}$ is neglected. This term plays a key role in describing relaxation and dissipation in an interacting many-electron system \cite{DAgosta2006Relaxation,DiVentra2007Stochastic}. Explicit expressions for $\mathbf{A}_{\mathrm{XC}}$ have, so far, become available only in the linear-response limit \cite{Vignale1996Currentdependent}. In view of the numerical complexity of the present simulation, we neglect this term. Relaxation phenomena occurring on longer time scales are, therefore, omitted from the outset.
 

We solve the time-dependent Kohn-Sham equations (Eq.~\ref{eq:tdks}) on a Cartesian grid with discretization $\sim 0.20$ a.u. in the laser polarization direction and $\sim 0.45$ a.u. perpendicular to the polarization direction in a cuboid cell of dimensions 9.28$\times$16.05$\times$10.21 a.u.$^3$ employing a nine-point stencil for the kinetic energy operator and a Bloch-momentum grid of $4^3$ $\mathbf k$ points. The time evolution is performed with a fourth-order Taylor approximation to the Hamiltonian with a time step of 0.02 a.u.~including a predictor-corrector step 
employing a pulse with a cosine-squared envelope, 
\begin{equation} \label{eq:Avec}
\mathbf A(t) = -\frac{\mathbf{F_0}}{\omega_{\mathrm{L}}} \cos(\omega_{\mathrm{L}} t + \phi_\mathrm{CE}) \left[ \cos( \frac{\pi}{2} \frac{t}{\tau_{\mathrm{p}}} )\right]^2
\end{equation}
where $\phi_{\mathrm{CE}}$ is the carrier-envelope phase (CEP) of the few-cycle infrared pulse of carrier frequency $\omega_{\mathrm{L}}$.


We analyze the time- and space-dependent microscopic current density 
\begin{equation}
\mathbf j(\mathbf r, t) = |e| \sum_i \frac{1}{2} \left[ \psi^*_i(\mathbf r, t) \left( -i  \mathbf \nabla + \mathbf A(t) \right) \psi_i(\mathbf r, t) + \mathrm{c.c.} \right] \quad .
\end{equation}
The macroscopic current density $J(t)$ along the laser polarization direction $\mathbf F_0$ is given by the average of $\mathbf j(\mathbf r, t)$ over the unit cell with volume $\Omega$
\begin{equation}
J(t) = \frac{1}{\Omega} \int_\Omega d\mathbf r \,\, \mathbf j(\mathbf r,t) \cdot \mathbf{F_0}/|F_0|
\end{equation}
and the corresponding polarization density is $P(t)= \int_{-\infty}^t J(t') dt'$. $P(t)$ gives the charge transferred per unit area at time $t$. Accordingly, the macroscopic charge transferred by the laser pulse follows from $P(t > \tau_\mathrm{p})$ after the conclusion of the pulse as 
\begin{equation}
Q_\mathrm{L} = P(t > \tau_\mathrm{p}) \, \mathcal{A}_\mathrm{eff} \quad , 
\label{eqcharge}
\end{equation}
where $\mathcal{A}_\mathrm{eff}$ is the effective surface perpendicular to the laser polarization direction of the target illuminated by the laser.

\begin{figure}
\centerline{\epsfig{file=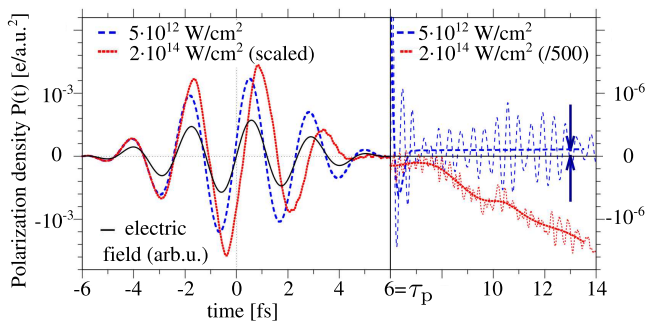,width=\columnwidth}}
\caption{Time-dependent macroscopic polarization density $P(t)$ along the laser polarization direction for two different laser intensities (blue dashed line $5\times10^{12}$ W/cm$^2$, red dotted line $2\times10^{14}$ W/cm$^2$, $P(t)$ scaled by the field amplitude ratio for comparison, black solid line laser field with photon energy 1.7 eV and pulse duration (full width) of $2 \tau_\mathrm{p}=$ 12 fs). Note the change in ordinate scale after the end of the pulse ($\tau_\mathrm{p} =$ 6 fs, right panel). Temporal averages (thick lines) over fast oscillations (thin lines). The vertical arrows indicate the persistent polarization after the laser pulse. }
\label{fig1}
\end{figure}

For a moderate laser intensity of $5\times 10^{12}$ W/cm$^2$ where the onset of the nonlinear response is expected, the time-dependent polarization density $P(t)$ (Fig.~\ref{fig1}) follows approximately adiabatically the applied electric force as expected within linear response. We note that because of the anisotropy of the crystal potential, the induced polarization vector can have a small component transverse to the axis of the laser polarization. After the pulse is over, the polarization density shows small-scale and fast oscillations. Their dominant oscillation frequency corresponds to the beating frequency between states in the valence and conduction band (period $\sim$0.5 fs). The average over these oscillations yields a small but finite sustained polarization density after the pulse, corresponding to a transferred charge density of about $1\times10^{-7}$ electrons per a.u.$^2$ (Fig.~1). This transferred charge remains, within the time interval covered by our propagation, constant to a good degree of approximation indicating the absence of any sustained current. The situation changes dramatically for an increased laser intensity of $2 \times10^{14}$ W/cm$^2$. During the laser pulse, the polarization density is distorted and phase shifted relative to the laser field. After the pulse has concluded at $t=\tau_\mathrm{p}$, the polarization density shows an almost linear decrease pointing to a constant current density flowing after the laser pulse is over.  Such a ballistic current will eventually relax due to dissipative processes such as electron-phonon coupling \cite{Franco2007Robust}, impurity, and disorder scattering on longer time scales ($\gtrsim 10$ fs) neglected in the present simulation. The appearance of quantum beats and a sustained current after the conclusion of the laser pulse qualitatively confirms earlier findings employing one-dimensional models \cite{Korbman2013Quantum,Kruchinin2013Theory}. 
The beating amplitude is drastically reduced mainly because of the three-dimensional rather than the one-dimensional density of states and because of the self-consistent inclusion of screening.

\begin{figure}
\centerline{\epsfig{file=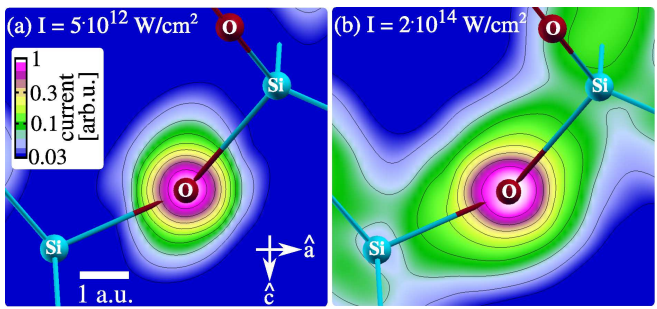,width= \columnwidth}}
\caption{Time-averaged current density $|\mathbf{j} (\mathbf{r},t>\tau_\mathrm{p})|$ in an $\hat a$-$\hat c$-plane of the trapezohedral SiO$_2$ lattice where the plane cuts through the O atom of a Si-O-Si bond, the laser is polarized in the $\hat c$ direction. (a) Laser intensity $5\times 10^{12}$ W/cm$^2$,  (b) laser intensity $2\times 10^{14}$ W/cm$^2$.}
\label{fig2}
\end{figure}

The time-averaged local current density after the laser pulse has concluded, $\mathbf{j}(\mathbf{r},t>\tau_\mathrm{p})$, gives first insight into the excitation mechanism. For low laser intensity (Fig.~\ref{fig2}(a)) it is centered around the O atoms with a slight elongation along the laser polarization direction (in Fig.~\ref{fig2}(a)) taken along the $\hat c$ axis of the crystal) while the current density near the Si atoms and in the interstitial region is negligible resembling localized atomiclike photoexcitation. At the higher laser intensity (Fig.~\ref{fig2}(b)) the situation is notably different: The current density distribution extends along the Si-O-Si bond axis and into the interstitial region near the Si atoms indicating laser-induced population of delocalized conduction band levels. 

\begin{figure}
\centerline{\epsfig{file=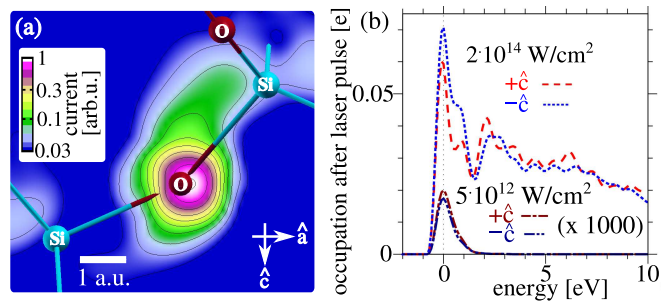,width= \columnwidth}}
\caption{(a) Snapshot of the current density for laser intensity $2\times 10^{14}$W/cm$^2$ taken near a laser field extremum at simulation time $t\approx -0.5$ fs in Fig.~\ref{fig1}. The electric field oriented along the $\hat c$-axis induces tunneling between neighboring atoms. (b) Occupation of conduction band states with positive (red lines) and negative (blue lines) $\mathbf k$-vectors with respect to the laser polarization direction $\hat c$ for a few-cycle pulse with laser intensities of $5\times 10^{12}$ W/cm$^2$ (lower graphs, magnified by a factor of 1000) and $2\times 10^{14}$ W/cm$^2$ (upper graphs). }
\label{fig3}
\end{figure}

The transition to the regime of a quasifree current can be visualized by snapshots of the time-dependent current density $\mathbf{j}(\mathbf{r},t)$ near the extremum of the laser field (Fig.~\ref{fig3}(a)). Within the strong field ionization model \cite{Keldysh1965} the excitation process is governed by the magnitude of the Keldysh parameter $\gamma = \omega \sqrt{2 \Delta}/ F_0$ with $\Delta$ the gap between the valence and conduction bands of the dielectric and $F_0$ the peak laser field strength. For $\gamma \gg 1$, multiphoton transition dominates while $\gamma \ll 1$ marks the regime of tunnel ionization. Accordingly, at an intensity of $2 \times 10^{14}$ W/cm$^2$ the Keldysh parameter( $\gamma \approx 0.7$) is in the tunneling excitation regime. For an isotropic static potential landscape, the tunneling current is expected to be oriented along the electric force direction exerted by the laser. In the present case of an anisotropic potential with the O-Si bonding direction at a finite angle relative to $\mathbf{F}_0$, the current displays a slight tilt (Fig.~\ref{fig3}(a)), consistent with the charge transfer to the Si atom and into the interstitial region. This directionality of the tunneling process in real space leaves its marks also in momentum space. For low laser intensities, coupling to the conduction band is weak and almost fully reversible. Moreover, $\mathbf{k}$ points oriented parallel and antiparallel to the laser amplitude are nearly equally populated after the laser pulse (Fig.~\ref{fig3}(b)) resulting in a vanishing free current. Transitions to the conduction band set in when the laser intensity surpasses the threshold for tunneling excitation. The latter can be estimated from the field strength $F_\mathrm{c}$ where the electrostatic potential difference between the O and the Si site (distance $d_\mathrm{O-Si} = $ 3.04 a.u.) reaches the order of magnitude of the excitation gap, $F_\mathrm{c} d_\mathrm{O-Si} \approx \Delta$. The resulting population of the  conduction band states after the conclusion of the laser pulse is orders of magnitude larger, and shows energy-dependent forward-backward asymmetries (Fig.~\ref{fig3}(b)).

\begin{figure}
\centerline{\epsfig{file=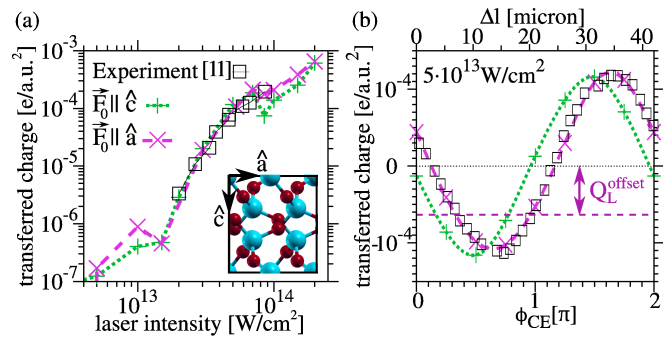,width=\columnwidth}}
\caption{Laser-induced charge transfer $Q_\mathrm{L}$ as a function of laser intensity in SiO$_2$ for photon energy 1.7 eV and pulse length 4.3 fs (FWHM intensity). Green crosses: laser polarized in $\hat c$ direction. Purple $X$: laser polarized in $\hat a$ direction. Black squares: experimental data \cite{Schiffrin2013Opticalfieldinduced}. (a) Intensity dependence of the $\phi_\mathrm{CE}$-maximized transferred charge, experimental data for an amorphous target. Inset: projection of the SiO$_2$ lattice structure onto the $\hat a$-$\hat c$-plane. (b) CEP dependence (see Eq.~\ref{eq:Avec}) at fixed intensity $5\times 10^{13}$ W/cm$^2$. Experimental data for a single crystal irradiated with polarization direction perpendicular to the $\hat c$ axis are plotted against the change $\Delta l$ in the propagation length through a fused silica wedge. Additional $\phi_\mathrm{CE}$-independent offset for polarization along $\hat a$ indicated by the horizontal dashed line.
\label{fig4}
}
\end{figure}

The present simulations can be compared with the first experimental data \cite{Schiffrin2013Opticalfieldinduced}. A comparison on an absolute scale for $Q_\mathrm{L}$ (Eq.~\ref{eqcharge}) would require the knowledge of the effective surface area $\mathcal{A}_\mathrm{eff}$ of the crystal illuminated by the laser with near-peak field strength and that effectively contributes to the charge transfer to nearby electrodes which is difficult to determine experimentally. We choose $\mathcal{A}_\mathrm{eff}$ to match the experimental value of $Q_\mathrm{L} = 0.6$ A fs at an intensity of  $5\times10^{13}$ W/cm$^2$ when integrating the current over $\sim 8$ fs before current damping would set in. Keeping the resulting scale factor $\mathcal{A}_\mathrm{eff}=8.7\times 10^{-14}$ m$^2$ 
fixed we find excellent agreement over more than two orders of magnitude for $Q_\mathrm{L}$ (Fig.~\ref{fig4}(a)) without any adjustable parameters. The steep rise clearly indicates the transition from a reversible non-linear bound polarization current to the excitation of a quasi-free current. 
We checked that this result (fig.~\ref{fig4}a) does not sensitively depend on the choice of the XC functional by also performing simulations employing the adiabatic local density approximation \cite{Perdew1981Selfinteraction}. We attribute the weak dependence on the accurate value of the band gap to the strongly nonlinear response beyond the lowest nonvanishing order of a multiphoton transition.


The experiment \cite{Schiffrin2013Opticalfieldinduced} has, furthermore, demonstrated that for a wave-form controlled few-cycle pulse, exquisite light-field control translates into control over the charge transfer. In particular, $Q_\mathrm{L}$ varies sinusoidally with the CEP of the few-cycle pulse (Fig.~\ref{fig4}(b)) clearly showing that the field amplitude, rather than the intensity is the parameter governing the charge transfer. We find excellent agreement with the experimental $\phi_\mathrm{CE}$ dependence for a laser intensity of $5\times 10^{13}$ W/cm$^2$ ($F_0=1.7$ V/\AA, Fig.~\ref{fig4}(b)). Our simulations predict a pronounced change of the CEP dependence with increasing intensity. Observation of the effect will require single crystals with well-defined orientation of the crystallographic axis relative to the laser polarization rather than fused silica targets \cite{Schiffrin2013Opticalfieldinduced}. 

We investigate the influence of the anisotropic electronic structure on the transferred charge by comparing simulations with laser polarization along the $\hat a$ and $\hat c$ axes of the crystal. We find that the dependence of $Q_\mathrm{L}$ on both the laser intensity and the CEP varies with orientation of laser polarization relative to the crystallographic axis in the high-field regime ($I \approx 10^{14}$ W/cm$^2$, Fig.~\ref{fig4}). Most notably, we observe a pronounced shift by $\approx \pi/4$ in $Q_\mathrm{L}$, $Q_\mathrm{L}(\hat a, \phi_\mathrm{CE}) \approx Q_\mathrm{L}(\hat c, \phi_\mathrm{CE}+ \pi/4)$ (Fig.~\ref{fig4}(b)). While the experiment \cite{Schiffrin2013Opticalfieldinduced} was designed to be sensitive only to the $\phi_\mathrm{CE}$-dependent part of the current and transferred charge, we find an additional unexpected $\phi_\mathrm{CE}$ independent contribution $Q^\mathrm{offset}_L$ to $Q_\mathrm{L}$ when the laser polarization is aligned along the $\hat a$ axis (Fig.~\ref{fig4}(b)). We trace its origin to the broken inversion symmetry along the $\hat a$ axis ($\hat a \to - \hat a$) of the SiO$_2$ crystal (see inset Fig.~\ref{fig4}(a)) leading to an average net charge transfer. Consequently, our simulation predicts that charge transfer in dielectrics induced by ultrafast pulses is possible even without the need for a CEP stabilized laser.


The present simulation provides a simple and transparent picture of the current and charge transfer dynamics. At low laser intensity, well within the linear-response regime ($I \le 10^{12}$ W/cm$^2$), neither a net current nor a charge displacement is induced. With increasing laser intensity, nonlinear effects become important. Starting from about $5\times 10^{12}$ W/cm$^2$, our simulations show that a finite amount of charge is transferred by nonlinear polarization currents during the laser pulse but no significant quasifree current flows after the pulse; i.e.~these polarization currents are almost completely reversible. Associating these currents with a field-induced AC conductivity $\sigma(\omega_\mathrm{L})$ at carrier frequency $\omega_\mathrm{L}$,
\begin{equation}
J(t) = \sigma(\omega_L) F(t)  \quad , 
\end{equation}
the nonlinear process of the charge displacement can be viewed as a reversible (sub) femtosecond-scale insulator to ``metal'' transition where $\sigma(\omega_\mathrm{L})$ increases by more than 20 orders of magnitude. The character of the field-induced currents changes significantly once the laser intensity is sufficiently high such that a substantial amount of electrons are nonadiabatically excited into the conduction band by tunneling excitation. The onset of a ballistic current in the material after the laser pulse is over is accompanied by a delocalized current density over the unit cell. This marks the precursor of dielectric breakdown for longer pulses. A finite conductivity, i.e.~a transition from a femtosecond ballistic current to a dissipative current will be established only on longer time scales ($\sim$20 fs as estimated from mobility data \cite{Williams1965Photoemission,Goodman1967Electron,Hughes1973ChargeCarrier}) by dissipative processes such as electron-phonon and defect scattering. In our simulation, the transition from nonlinear polarization to the regime where quasifree ballistic electron currents dominate occurs at a laser intensity of about $5\times 10^{13}$ W/cm$^2$. We find the amount of charge transferred is influenced by the intensity, pulse shape, and polarization direction of the laser pulse, indicating that the charge separation depends sensitively on the details of the potential landscape and bond structure.

The present results suggest opportunities for future investigations of the nonequilibrium electron dynamics on the femtosecond scale, in particular the transition regime from ballistic to dissipative electron transport in a pump-probe setting. This would require the inclusion of dissipation and quantum transport beyond the ballistic limit, e.g.~through nonadiabatic exchange-correlation functionals within an open-quantum system approach \cite{Wijewardane2005Timedependent,DAgosta2006Relaxation}.

Helpful discussions with V. Yakovlev, M. Ivanov, and M. Stockman are gratefully acknowledged. This work was supported by the FWF (Austria), SFB-041 ViCoM, SFB-049 Next Lite, and P21141-N16. G.W.\ thanks the IMPRS-APS for financial support. X.-M.T.\ was supported by a Grant-in-Aid for Scientific Research (Grant No.~C24540421) from the JSPS. K.Y.\ acknowledges support by the Grants-in-Aid for Scientific Research Grants No.\ 23340113 and No.~25104702. Calculations were performed using the Vienna Scientific Cluster (VSC), the supercomputer at the  Institute of Solid State Physics, University of Tokyo, and the T2K-Tsukuba at the Center for Computational Sciences, University of Tsukuba.

\section*{References}
\bibliography{bibliography_etal}
\end{document}